\documentclass{aa}
\usepackage{epsfig}

\def\etal{{\rm et al.\thinspace}}
\def\eg{{\rm e.g.}}

\def\ie{{\rm i.e.\ }}
\def\cf{{\rm cf.\ }}

\def\spose#1{\hbox to 0pt{#1\hss}}
\def\ltsimm{\mathrel{\spose{\lower 3pt\hbox{$\sim$}}
	\raise 2.0pt\hbox{$<$}}}
\def\gtsimm{\mathrel{\spose{\lower 3pt\hbox{$\sim$}}
	\raise 2.0pt\hbox{$>$}}}
\def\Mdot{\hbox{${\dot M}$} \,}

\def\km{{\rm\thinspace km}}
\def\cm{{\rm\thinspace cm}}
\def\pc{{\rm\thinspace pc}}
\def\s{{\rm\thinspace s}}
\def\yr{{\rm\thinspace yr}}
\def\g{{\rm\thinspace g}}

\def\kmps{\hbox{${\rm\km\s^{-1}\,}$}}
\def\erg{{\rm\thinspace erg}}

\def\Msol{\hbox{${\rm\thinspace M_{\odot}}$}}
\def\Msolpyr{\hbox{${\rm\Msol\yr^{-1}\,}$}}
\def\ergpcm2ps{\hbox{${\rm\erg\cm^{-2}\s^{-1}\,}$}}
\def\pcm3{\hbox{${\rm\cm^{-3}\,}$}}
\def\gpcm3{\hbox{${\rm\g\cm^{-3}\,}$}}
\def\gpcm3ps{\hbox{${\rm\g\cm^{-3}\s^{-1}\,}$}}
\def\gps{\hbox{${\rm\g\s^{-1}\,}$}}

\begin{document}
   
\title{The Thermal Stability of Mass-Loaded Flows}

\author{J.M. Pittard, T.W. Hartquist, I. Ashmore}

\institute{Department of Physics and Astronomy, The University of Leeds, 
        Woodhouse Lane, Leeds, LS2 9JT, UK
\\}

\offprints{J. M. Pittard, \email{jmp@ast.leeds.ac.uk}}

\date{Received date / Accepted date}

\abstract{We present a linear stability analysis of a flow undergoing
conductively-driven mass-loading from embedded clouds. We find that 
mass-loading damps isobaric and isentropic
perturbations, and in this regard is similar to the effect of thermal 
conduction, but is much more pronounced where many embedded clumps exist. 
The stabilizing influence of mass-loading is wavelength
independent against isobaric (condensing) perturbations, but wavelength
dependent against isentropic (wave-like) perturbations. We derive equations
for the degree of mass-loading needed to stabilize such perturbations.
We have also made 1D numerical simulations of a mass-loaded radiative 
shock and demonstrated the damping of the overstability when mass-loading
is rapid enough.
}

\titlerunning{The Thermal Stability of Mass-Loaded Flows}
\authorrunning{Pittard, Hartquist \& Ashmore}

\maketitle

\label{firstpage}

\section{Introduction}
\label{sec:intro}

The thermal instability of radiative media was examined by
Field (\cite{F1965}), who considered both the presence and absence of 
thermal conduction, and derived the growth rates of isobaric and isentropic
perturbations. The first numerical calculations of catastrophic cooling in
shock heated gas were performed by Falle (\cite{F1975}, \cite{F1981}).
Radiative shocks were shown to exhibit a global overstability
by Langer \etal (\cite{L1981}), and have since been extensively examined
(\eg, Chevalier \& Imamura \cite{CI1982}; Imamura, Wolff \& Duisen 
\cite{IWD1984}; Gaetz, Edgar \& Chevalier \cite{GEC1988}; Blondin \& Cioffi
\cite{BC1989}; Strickland \& Blondin \cite{SB1995}; Walder \& Folini
\cite{WF1996}).

The interaction of radiative flows with cold embedded clouds is known to
significantly modify these flows (see, \eg, Pittard, Dyson \& Hartquist
\cite{PDH2001}, and references therein). However, there has yet to be
an investigation into how mass-loading may affect the {\em stability} 
properties of such flows. This is the aim of this paper. 

In Sec.~\ref{sec:lin_stab} we perform a linear stability analysis of
a static medium, in thermal equilibrium, undergoing conductively-driven 
mass-loading. In Sec.~\ref{sec:numerics} we present numerical models of
mass-loaded radiative shocks to examine the suppression of thermal instability
in them. We finish in Sec.~\ref{sec:discuss} with a discussion on the 
conditions necessary for mass-loading to suppress the thermal instability
in a typical planetary nebula, and provide numerical estimates for the
Helix nebula.

\section{Instability in a uniform medium, including the effects of
conductively-driven mass-loading}
\label{sec:lin_stab}
The dynamics are governed by the standard hydrodynamic equations for
plane parallel flow,

\begin{equation}
\label{eq:continuity}
\frac{\partial \rho}{\partial t} + \frac{\partial(\rho v)}{\partial z} =
q_{0}\left(\frac{T}{T_{0}}\right)^{\lambda} - q_{0},
\end{equation}

\begin{equation}
\label{eq:mtm}
\frac{\partial (\rho v)}{\partial t} + \frac{\partial(\rho v^{2})}{\partial z} 
 + \frac{\partial P}{\partial z} = 0,
\end{equation}

\begin{eqnarray}
\label{eq:energy}
\lefteqn{\frac{1}{2}\frac{\partial (\rho v^{2})}{\partial t}  + 
\frac{1}{(\gamma-1)}\frac{\partial P}{\partial t}} \nonumber \\ & &
\hspace{20mm} + \frac{\partial}{\partial z} \left(\frac{1}{2}\rho v^{3} + 
\frac{\gamma}{(\gamma-1)} vP\right) = - \mathcal{L} \rho,
\end{eqnarray}

\noindent together with an equation of state,

\begin{equation}
\label{eq:equn_state}
P = \frac{R}{\mu} \rho T.
\end{equation}

\noindent We have assumed that the clouds are all at their equilibrium radius
so that if $T < T_{0}$ they grow and that if $T > T_{0}$ they evaporate
(McKee \& Cowie \cite{MC1977}). 
$q_{0}$ has the dimensions of a mass-loading rate per unit volume.

The equilibrium state is characterized by $\rho = \rho_{0}$, $T = T_{0}$,
$v = 0$, and $\mathcal{L}(\rho_{0},T_{0}) = 0$. Assuming perturbations of the
form

\begin{equation}
\label{eq:perturb}
a(z,t) = a_{1} \; {\rm exp} \;(nt + ikz),
\end{equation}

\noindent we find the linearized equations for the perturbations to be

\begin{equation}
\label{eq:lin_cont}
n \rho_{1} + \rho_{0} ikv_{1} = q_{0} \lambda T_{1}/T_{0},
\end{equation}

\begin{equation}
\label{eq:lin_mtm}
n \rho_{0} v_{1} + ikP_{1} = 0,
\end{equation}

\begin{equation}
\label{eq:lin_energy}
\frac{n}{\gamma -1} + \frac{\gamma}{\gamma - 1} ikv_{1}P_{0} = 
-\rho_{0} \mathcal{L}_{T} T_{1} - \rho_{0}\mathcal{L}_{\rho} \rho_{1},
\end{equation}

\noindent and

\begin{equation}
\label{eq:lin_state}
P_{1} - \frac{R}{\mu} \rho_{1} T_{0} - \frac{R}{\mu} \rho_{0} T_{1} = 0,
\end{equation}

\noindent where $\mathcal{L}_{\rho} \equiv 
(\partial \mathcal{L}/\partial \rho)_{T}$ and $\mathcal{L}_{T} \equiv 
(\partial \mathcal{L}/\partial T)_{\rho}$ are evaluated for the
equilibrium state. We are left with 4 variables ($\rho_{1}, \;P_{1}, \;T_{1}$,
and $v_{1}$) in 4 equations. The resulting dispersion relation is

\begin{eqnarray}
\label{eq:dispersion}
\lefteqn{n^{3} + n^{2} c \left(k_{T} + \frac{k_{m}}{\gamma}\right) +
n c^{2} \left(k^{2} + \frac{k_{\rho} k_{m}}{\gamma}\right)} \nonumber \\ & &
\hspace{30mm} + \frac{c^{3} k^{2}}{\gamma} (k_{T} - k_{\rho} + k_{m}) = 0.
\end{eqnarray}

\noindent The adiabatic speed of sound is 
$c = (\gamma P_{0}/\rho_{0})^{1/2}$, and we have introduced the
wavenumbers

\begin{equation}
\label{eq:k_rho}
k_{\rho} = \frac{\mu (\gamma -1) \rho_{0} \mathcal{L}_{\rho}}{R c T_{0}},
\end{equation}

\begin{equation}
\label{eq:k_T}
k_{T} = \frac{\mu (\gamma -1) \mathcal{L}_{T}}{R c},
\end{equation}

\noindent and

\begin{equation}
\label{eq:k_m}
k_{m} = \frac{\gamma \lambda q_{0}}{\rho_{0} c}.
\end{equation}


By introducing the non-dimensional variables

\begin{eqnarray}
\label{eq:nd_var}
\lefteqn{y = \frac{n}{kc}, \hspace{5mm} \sigma_{\rho} = \frac{k_{\rho}}{k}, 
\hspace{5mm} \sigma_{T} = \frac{k_{T}}{k},} \hspace{40mm} \nonumber \\ 
\sigma_{m} = \frac{k_{m}}{k}, &
\hspace{5mm} \sigma_{T}' = \sigma_{T} + \sigma_{m}, \hspace{10mm}
\end{eqnarray}

\noindent we can write the dispersion equation in the form

\begin{equation}
\label{eq:disp2}
y^{3} + y^{2}\sigma_{T}' + y(1 + \sigma_{\rho}{\sigma_m}) +
(\sigma_{T} + \sigma_{m} \gamma - \sigma_{\rho})/\gamma = 0.
\end{equation}

\noindent The coefficient in the y term can be removed by the introduction
of the variable

\begin{equation}
\label{eq:ydash}
y' = \frac{y}{(1 + \sigma_{\rho}\sigma_{m})^{1/2}}.
\end{equation}
 
\noindent The dispersion relation then becomes

\begin{equation}
\label{eq:disp3}
y'^{3} + \frac{\sigma_{T}'}{(1 + \sigma_{\rho}{\sigma_m})^{1/2}}y'^{2} +
y' + \frac{\sigma_{T} + \sigma_{m} \gamma - \sigma_{\rho}}{\gamma
(1 + \sigma_{\rho}{\sigma_m})^{3/2}} = 0.
\end{equation}

\noindent This equation is now in the same form as Eq.~18 in 
Field (\cite{F1965}). 

The growth of the isobaric instability (which Field refers to as
a condensation mode) requires

\begin{equation}
\label{eq:isobaric}
\sigma_{T} - \sigma_{\rho} < -\gamma \sigma_{m}.
\end{equation}

\noindent Since $\sigma_{m}$ is positive by definition, mass-loading always
acts to reduce the growth rate of this instability mode (as Field found for
conduction). However, unlike the corresponding equation including 
conduction, Eq.~\ref{eq:isobaric}
is independent of $k$, so mass-loading stabilizes all wavelengths equally
effectively against isobaric perturbations. Rewriting Eq.~\ref{eq:isobaric},
we find that the instability is suppressed if

\begin{equation}
\label{eq:isobaric_sup}
\frac{q_{0}}{\rho_{0}} > \frac{(\gamma - 1) \mu}{\gamma \lambda R}
\left(\frac{\rho_{0} \mathcal{L}_{\rho}}{T_{0}} - \mathcal{L}_{T}\right).
\end{equation}

\noindent This is roughly equivalent to requiring 
that the mass-loading timescale be less than the cooling timescale of the
medium.

The growth of the isentropic instability requires

\begin{equation}
\label{eq:isentropic}
\sigma_{T} + \frac{\sigma_{\rho}}{\gamma - 1} < -\frac{\gamma}{\gamma-1} 
\sigma_{m} \sigma_{\rho} (\sigma_{T} + \sigma_{m}).
\end{equation}

\noindent Mass-loading is again always stabilizing, but is not equally
effective at all wavelengths in suppressing isentropic perturbations. 
Rewriting Eq.~\ref{eq:isentropic} we find that the
critical wavenumber above which perturbations are stabilized is

\begin{equation}
\label{eq:crit_k}
k_{\rm c} = \sqrt{- \frac{k_{m} k_{\rho} (k_{T} + k_{\rho})}{(\gamma - 1)
k_{T} + k_{\rho}}}.
\end{equation}

\noindent and that the isentropic instability is suppressed if

\begin{equation}
\label{eq:isentropic_sup}
q_{0} > -\frac{k^{2} c^{4}}{\gamma^{2} (\gamma - 1) \lambda 
\mathcal{L}_{\rho}} \frac{(\gamma - 1)\mathcal{L}_{T} +
\rho_{0}\mathcal{L}_{\rho}/T_{0}}{\mathcal{L}_{T} + 
\rho_{0}\mathcal{L}_{\rho}/T_{0}}.
\end{equation}

Over the temperature range $5 \times 10^{5} \ltsimm T \ltsimm 
5 \times 10^{7}\;{\rm K}$, a good approximation is 
$\mathcal{L} = \rho \Lambda T^{-1/2}$ (\eg, Kahn \cite{K1976}). This leads to
greatly simplified versions of Eqs.~\ref{eq:isobaric_sup} 
and~\ref{eq:isentropic_sup}, which respectively become

\begin{equation}
\label{eq:isobaric_sup2}
\frac{q_{0}}{\rho_{0}} > \frac{3}{2} \frac{(\gamma - 1) \mu}{\gamma \lambda R}
\rho_{0} \Lambda T_{0}^{-3/2},
\end{equation}

\noindent and

\begin{equation}
\label{eq:isentropic_sup2}
q_{0} > \frac{2 k^{2} c^{4}}{\gamma^{2} (\gamma - 1) \lambda 
\Lambda T_{0}^{-1/2}} \left[1 - \frac{1}{2}(\gamma - 1)\right],
\end{equation}

\noindent and to the relationship
$k_{\rho}/k_{T} = \sigma_{\rho}/\sigma_{T} -2$.

\section{Hydrodynamical calculations}
\label{sec:numerics}

The nature of the overstability of radiative shocks is known to depend on
the temperature dependence of the local cooling rate. For a power-law
dependence (\eg, $\mathcal{L} \rho \propto T^{\alpha}$), the system is
overstable for values of $\alpha$ below some critical value, 
$\alpha_{\rm cr}$. Systems with $\alpha \gtsimm \alpha_{\rm cr}$ are
stable. Previous numerical work has shown that $\alpha_{\rm cr} \approx 0.4$ 
(\eg, Imamura \etal \cite{IWD1984}; Strickland \& Blondin \cite{SB1995}),
in good agreement with the linear stability analysis of 
Chevalier \& Imamura (\cite{CI1982}).

\begin{figure}[t]
\begin{center}
\hspace{7mm}
\vspace{10mm}
\psfig{figure=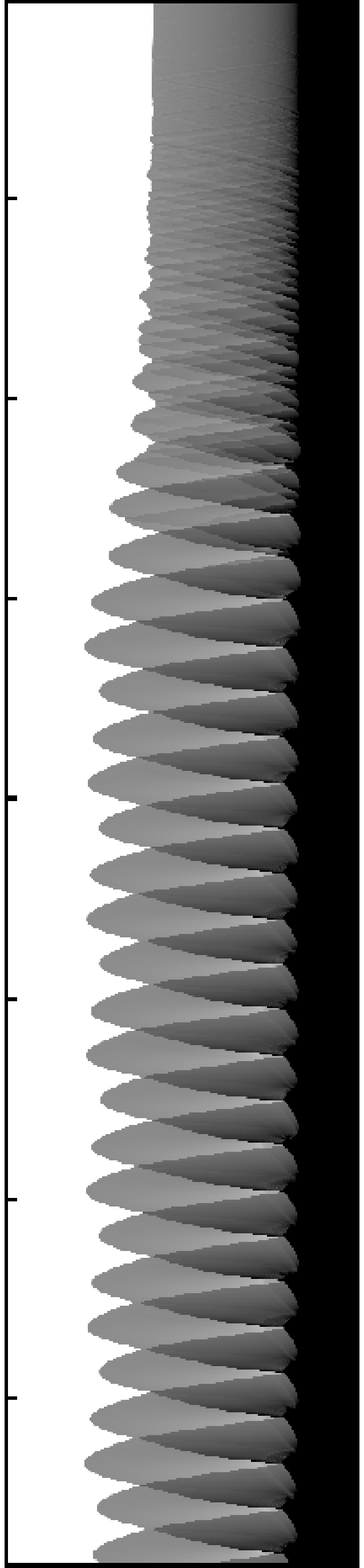,width=3.5cm}
\psfig{figure=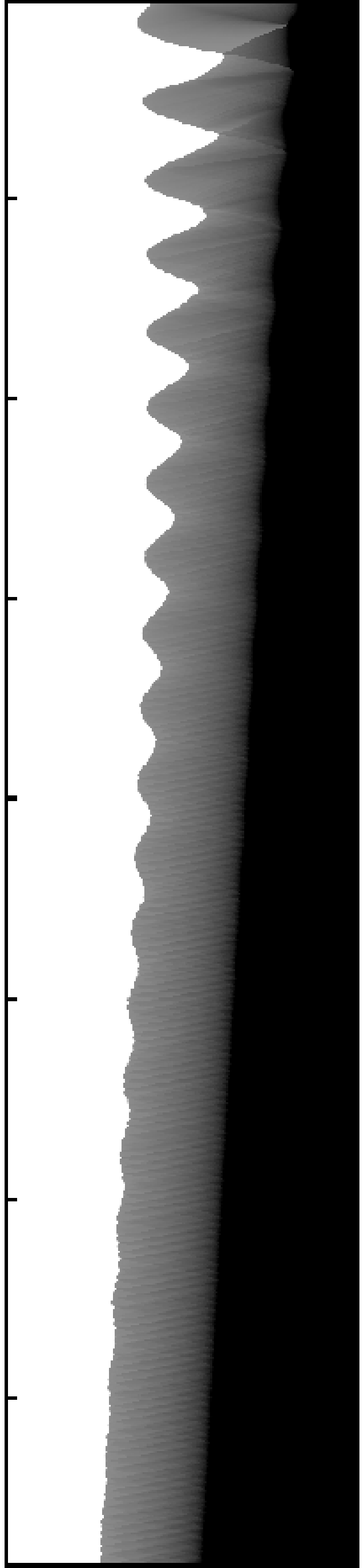,width=3.5cm}
\end{center}
\caption[]{Spacetime diagrams of a radiative shock for $M=10$, 
$\alpha = 0.0$ (\cf Fig.~3 in Strickland \& Blondin \cite{SB1995}). 
The tick marks on the
vertical axis are in units of $25 L_{\rm c}/v_{0}$ where $L_{\rm c}$ is the 
cooling length of the shock (see Strickland \& Blondin \cite{SB1995}).
The width of the horizontal axis is $1.88 L_{\rm c}$. The supersonic flow
enters the grid from the left, and the cooled postshock gas flows off the grid
to the right. The grayscale shows the density of the gas, with
lighter shades signifying diffuse gas and darker regions denoting 
higher density. In the left panel, $q_{0} = 0$, and the system exhibits 
linear growth for the first $\sim 40\%$ of the run. In the right panel, 
where $q_{0} = 150$, the initial conditions are far from
equilibrium for a mass-loaded shock, and the system exhibits powerful 
oscillations, though these are quickly damped as expected from the 
linear stability analysis in Sec.~\ref{sec:lin_stab}.}
\begin{center}
\setlength{\unitlength}{1mm}
\begin{picture}(110,50)(0,0)
\put(5,237){\makebox(0,0){\Large time}}
\linethickness{1mm}
\put(5,202){\line(0,1){30}}
\linethickness{10mm}
\put(5,202){\line(-1,1){4}}
\put(5,201.75){\line(-1,1){4}}
\put(5,201.5){\line(-1,1){4}}
\put(5,202){\line(1,1){4}}
\put(5,201.75){\line(1,1){4}}
\put(5,201.5){\line(1,1){4}}
\put(45,122){\makebox(0,0){\Large flow direction}}
\linethickness{1mm}
\put(65,117){\line(-1,0){40}}
\linethickness{2mm}
\put(65,117){\line(-1,1){4}}
\put(65.25,117){\line(-1,1){4}}
\put(65.5,117){\line(-1,1){4}}
\put(65,117){\line(-1,-1){4}}
\put(65.25,117){\line(-1,-1){4}}
\put(65.5,117){\line(-1,-1){4}}
\end{picture}
\end{center}
\label{fig:m10}
\end{figure}

To investigate the effect of mass-loading on the stability of a 
radiative shock, we have computed 1D numerical calculations 
for a Mach 10 shock with the condition that $\alpha = 0.0$. 
Our computations were initialized in a similar fashion to that
presented by Strickland \& Blondin (\cite{SB1995}) for the case of an 
isolated planar shock, and were performed using the same hydrodynamical
code, VH-1 (see Blondin \etal \cite{BKFT1990}).  

Briefly, we assumed a 
steady-state radiative shock in the centre of the grid, with pre-shock flow
from the left grid boundary, and cold, dense, post-shock material 
exiting the right grid boundary. The most common downstream condition for
similar calculations in the literature is that of a reflecting wall, but
a continuous outflow has several benefits. For instance, feedback between
the cold dense post-shock layer and the cooling gas is properly treated, 
and the problem more closely resembles the common occurence of an 
isolated shock in the interstellar medium. We also repeat the cautionary 
note of Strickland \& Blondin (\cite{SB1995}) that the boundary conditions
play an important role in determining the stability of the shock. Therefore,
it is desirable to place the shock well away from grid boundaries.
We refer the reader to Strickland \& Blondin (\cite{SB1995}) for
a fuller description of the code and initial conditions. In our runs
we assumed $\gamma = 5/3$.

In the left panel of Fig.~\ref{fig:m10} 
we show the development of the overstability
for a Mach 10 flow with $\alpha = 0.0$ and $q_{0} = 0.0$ 
(\ie no mass-loading). The pre-shock density and velocity are set to
unity. At the beginning of the simulation, corresponding to the
top of the plot, the flow is initialized as closely as possible to the 
steady state solution for a radiative shock with no mass-loading. 
The overstability is excited from weak
perturbations produced by the numerical mapping of the steady state solution
onto the computational grid. Linear growth was observed for about 
the first 40\% of the run, after which the amplitude of the oscillations
saturate and the system enters the nonlinear regime. 
The behaviour 
shown in the left panel of Fig.~\ref{fig:m10} is
in excellent agreement with Fig.~3 of 
Strickland \& Blondin (\cite{SB1995}) where the overstability for
a Mach 40 flow is displayed.
 
In the right panel of Fig.~\ref{fig:m10} 
we show the evolution of a mass-loaded 
radiative shock with $q_{0} = 150.0$ and $\lambda = 5/2$. 
The process of mass-loading increases the density and thermal pressure, and 
decreases the temperature and velocity of the flow. The pressure
increase causes the shock position to initially expand away from the
dense cooling layer. However, the enhanced density leads to more 
rapid cooling, and thermal pressure support is soon lost. This 
causes the shock to reverse its direction of motion and to fall
towards the cold dense layer. The shocked gas is then 
repressurized and this cycle is repeated. With each cycle, the 
amplitude of the oscillations decreases as the mass-loading damps this
instability.

With $\alpha = 0.0$, $\mathcal{L} = 1$, $\mu/{\rm R} = 1$ and the 
post-shock values $\rho = 4$, $T = 0.193$, Eq.~\ref{eq:isobaric_sup} 
shows that we require $q_{0} > 13$ for mass-loading to begin to
damp the overstability. In practice, we find that we need values
appreciably larger than this for strong damping, although the oscillation
amplitude does decrease slightly for values of $q_{0}$ near 13. 
There are a couple of reasons why we need higher values of $q_{0}$ than
suggested by Eq.~\ref{eq:isobaric_sup}. The principal reason is due
to the fact that adding mass to the shocked gas decreases the cooling
timescale while simultaneously increasing the mass-loading timescale.
Hence, while the initial addition of mass to the immediate post-shock
flow may be at a rate such that the mass-loading timescale is less 
than the cooling timescale, the difference between the two timescales
will diminish with time and may even be reversed. 

Secondly, the large dynamic range in temperature which exists for
high Mach number shocks (\eg, for a Mach number of 40, the immediate 
post-shock temperature is almost 3 orders of magnitude greater than the 
ambient temperature) leads to a smaller value of $T_{\rm avg}/T_{\rm ps}$
than is the case for lower Mach number shocks. Here $T_{\rm avg}$ is 
the average temperature over the cooling length of the shock,
and $T_{\rm ps}$ is the immediate post-shock temperature.  
We find that as we reduce the
shock Mach number, the value of $q_{0}$ needed for significant damping 
more closely matches that from Eq.~\ref{eq:isobaric_sup}.

While we have not been concerned with the ionization state of 
the gas in this model, it is worth noting that the
radiative cooling rate can be greatly enhanced when neutral gas is
introduced into hot plasma, and that the ionization energy can also be
significant under such conditions (Slavin \etal \cite{SSB1993}).

\section{Discussion}
\label{sec:discuss}
As a relevant example, the above results can be applied to 
planetary nebulae (PNe). Such objects often display clumps which
appear to mass-load the shocked wind of the central star. The
clumps were presumably formed by instabilities in the atmospheres of
the red supergiant stars, prior to the evolution of the central star
into a hot object with a fast wind. 
In the following we derive relationships for the number density of
clouds and their total mass, with the condition that mass injection
from the clouds is able to suppress the thermal instability.

Since the central stars of PNe appear to fall into 3 groups (\eg,
Kudritzki \etal \cite{K1997}), with winds spanning a large range of
mass-loss rate ($\Mdot \sim 10^{-9} - 10^{-6} \Msolpyr$), we choose not
to focus on an individual nebula. Instead, we note that the typical
thermal pressure in PNe, wind-blown-bubbles, and starburst superwinds
is comparable to the pressure in clumps embedded within them, 
$P/k \sim 10^{7}\;{\rm cm^{-3}\;K}$. In PNe, this must balance the 
ram pressure from the wind of the central star. The post-shock
number density is then $n = 4 \times 10^{-7} v_{7}^{-2} k/\mu (P/k)_{7} = 
55 v_{7}^{-2} (P/k)_{7} \;\pcm3$, where $v_{7}$ is the wind speed in units 
of $1000 \kmps$ and $(P/k)_{7}$ is $P/k$ in units of 
$10^{7}\;{\rm cm^{-3}\;K}$.
In the last step, and throughout the remainder of these calculations,
we set $\mu = 10^{-24} \;{\rm g}$. The post-shock temperature, 
$T = 1.36 \times 10^{5} v_{7}^{2} \;{\rm K}$.  

To suppress the thermal instability by mass-loading, 
Eq.~\ref{eq:isobaric_sup} gives
$q_{0} \gtsimm 10^{-32} v_{7}^{-7} (P/k)_{7}^{2} \;\gpcm3ps$, 
where we have adopted $\Lambda = 1.33 \times 10^{-19}/\mu^{2}$ 
and $\alpha = -{1\over2}$ (\cf Kahn \cite{K1976}), and 
substituted for $T$ and $n$.
The mass evaporation rate from a single clump is
$\dot{m} = 2.75 \times 10^{4} T^{5/2} R_{\rm pc} (30/{\rm ln}\Lambda)
\;\gps$ (Cowie \& McKee \cite{CM1977}), where $T$ is the temperature
of the ambient surroundings, and $R_{\rm pc}$ is the radius of the
clump in parsecs. Since ${\rm ln}\Lambda \sim 30$,
$\dot{m} \sim 1.9 \times 10^{17} v_{7}^{5} R_{\rm pc} \;\gps$, and 
the required number density of clumps is $n_{\rm c} = q_{0}/\dot{m}
\gtsimm 1.6 \times 10^{6} v_{7}^{-12} R_{\rm pc}^{-1} (P/k)_{7}^{2} 
\;{\rm pc^{-3}}$.

Observations of the Helix nebula (NGC~7293) indicate that the 
density within the clumps, $N_{\rm c} \sim 10^{6} \pcm3$, while they
are unresolved on spatial scales of $\sim 10^{-3} \pc$ 
(O'Dell \& Handron \cite{DH1996}). Adopting 
$R_{\rm pc} = 10^{-3}$, $v_{7} = 2$, and $(P/k)_{7} = 1$ yields a number 
density of clumps of $n_{\rm c} \gtsimm 4 \times 10^{5} \;{\rm pc^{-3}}$. 
Since the radius of the Helix nebula is $\sim 0.2 \;\pc$, we then 
expect $\sim 1600$ clumps, with a total mass of $\sim 0.1 \;\Msol$. 
While these estimates are in 
good agreement with observational inferences from the Helix nebula, 
small changes in $v_{7}$ can greatly affect these values.
For the evaporation to be smooth on a spatial scale of $1/k_{\rm m}$, we
require that the clumps are smaller than $L = 2\pi/k_{\rm m}$, \ie 
$L \ltsimm 1.5 \times 10^{-2} v_{7}^{6} (P/k)_{7}^{-1} \pc$.
With the above parameters, $L \ltsimm 1 \pc$, and is consistent with
the derived number density of clumps.


\begin{acknowledgements}
We would like to thank the referee, John Raymond, for timely and 
constructive comments.
JMP would also like to thank PPARC for the funding of a PDRA position. 
This research has made use of NASA's Astrophysics Data System Abstract 
Service. 
\end{acknowledgements}

\end{document}